\documentclass[runningheads]{llncs}
\usepackage[T1]{fontenc}
\usepackage{cite}
\usepackage{amsmath,amssymb,amsfonts}
\usepackage{algorithmic}
\usepackage{graphicx}
\usepackage{textcomp}
\usepackage{xcolor}
\usepackage{multirow}
\usepackage{multicol}
\usepackage{booktabs}
\usepackage{amsfonts}
\usepackage{graphicx}
\begin{document}
\title{Voice Conversion with Denoising Diffusion Probabilistic GAN Models}
\author{ Xulong Zhang, Jianzong Wang\thanks{Corresponding author: Jianzong Wang, jzwang@188.com.},  Ning Cheng, Jing Xiao}
\authorrunning{Zhang et al.}
\institute{\textit{Ping An Technology (Shenzhen) Co., Ltd.}}
\maketitle              % typeset the header of the contribution
\begin{abstract}
Voice conversion is a method that allows for the transformation of speaking style while maintaining the integrity of linguistic information. There are many researchers using deep generative models for voice conversion tasks. Generative Adversarial Networks (GANs) can quickly generate high-quality samples, but the generated samples lack diversity. The samples generated by the Denoising Diffusion Probabilistic Models (DDPMs) are better than GANs in terms of mode coverage and sample diversity. But the DDPMs have high computational costs and the inference speed is slower than GANs. In order to make GANs and DDPMs more practical we proposes DiffGAN-VC, a variant of GANs and DDPMS, to achieve non-parallel many-to-many voice conversion (VC). We use large steps to achieve denoising, and also introduce a multimodal conditional GANs to model the denoising diffusion generative adversarial network. According to both objective and subjective evaluation experiments, DiffGAN-VC has been shown to achieve high voice quality on non-parallel data sets. Compared with the CycleGAN-VC method, DiffGAN-VC achieves speaker similarity, naturalness and higher sound quality.

\end{abstract}
\section{Introduction}
Voice conversion (VC), a branch of speech signal processing also referred to as speech style transfer, entails the transformation of the linguistic attributes of the source speaker to those of the target speaker, while preserving the linguistic content of the input voice. This technology pertains to modify the linguistic style of the speech samples. VC can be viewed as a regression problem whose goal is to build a mapping function between the speech features of the target and source speaker. VC technology has important applications in many fields, including privacy and identity protection, speech imitation and camouflage, and speech enhancement \cite{DBLP:journals/taslp/TodaNS12,tang2023learning,tang2023vq,si2022boosting}. This technology is also capable of transforming standard reading speech into stylized speech, including emotional and falsetto speech~\cite{DBLP:journals/taslp/TodaNS12,tang2023emomix,tang2023qi}. At the same time, it can be used in music for the conversion of singers' vocal skills \cite{DBLP:conf/icassp/DengYLW020}. It also can be used to help people with language disorders voice assistance \cite{DBLP:journals/taslp/TodaNS12,DBLP:journals/speech/NakamuraTSS12}, and can convert voice which contains noise in meeting to clear voice \cite{DBLP:conf/interspeech/KanekoKHK17}.

Traditional voice conversion technology is developed based on parallel data, which comprises recordings of both the target and source speakers speaking identical content. The current models using this data mainly include (1) statistical methods (GMMs) \cite{DBLP:journals/taslp/TodaBT07}; (2) exemplar-based methods (NMF) \cite{DBLP:journals/taslp/WuVCL14}; (3) Neural network-based methods, such as feedforward neural networks (FNNs) \cite{DBLP:conf/slt/MohammadiK14}, recurrent neural networks (RNNs) \cite{DBLP:conf/icassp/SunKLM15}, convolutional neural networks (CNNs) \cite{DBLP:conf/interspeech/KanekoKHK17}. However, we often lack parallel data sets to train model. Simultaneously, collecting and producing a large parallel data set is very laborious. In addition, even if we have collected the corresponding data, we also need to carry out time alignment operation \cite{DBLP:conf/interspeech/HelanderSNSG08}. Many researchers have also explored non-parallel VC methods that don’t need parallel data. Due to unfavorable training conditions, it is not as good as parallel VC methods in terms of conversion effect and speech quality. Therefore, this is a challenging and important task. This paper focuses on the development of a non-parallel VC method with the same high audio quality and conversion effects as parallel methods.

In order to implement non-parallel VC using voice data, probabilistic depth generation models have been introduced in many studies recently, mainly including RBM based methods, variational autoencoders (VAEs) \cite{DBLP:journals/corr/KingmaW13, DBLP:conf/icml/QianZCYH19, DBLP:conf/interspeech/DingG19, DBLP:journals/tetci/HuangLHLPTW20, DBLP:conf/icassp/QianJHM20, DBLP:journals/taslp/KameokaKTH19, DBLP:conf/interspeech/TobingWHKT19} based methods and generative adversarial networks (GANs) \cite{DBLP:journals/cacm/GoodfellowPMXWO20} based methods. GANs is a powerful generative model that can learn the generation distribution. It has shown great success in the fields of image style transfer, image quality improvement and image generation. Among them, CycleGAN-VCs~\cite{DBLP:conf/eusipco/KanekoK18,DBLP:conf/icassp/KanekoKTH19,DBLP:conf/interspeech/KanekoKTH20,DBLP:conf/icassp/KanekoKTH21,DBLP:conf/slt/KameokaKTH18,DBLP:conf/interspeech/KanekoKTH19} have attracted extensive attention from researchers.

Recently, Denoising Diffusion Probabilistic Models (DDPM) \cite{DBLP:conf/nips/HoJA20, DBLP:conf/icassp/LuWWRYT22} shows good performance in various generative tasks such as image generation \cite{DBLP:conf/cvpr/LugmayrDRYTG22, DBLP:conf/siggraph/SahariaCCLHSF022}, neural acoustic coding \cite{DBLP:conf/iclr/KongPHZC21, DBLP:conf/iclr/ChenZZWNC21}, speech enhancement\cite{DBLP:conf/icassp/LuWWRYT22}, and speech synthesis \cite{DBLP:conf/aaai/Liu00CZ22, DBLP:conf/interspeech/JeongKCCK21}. Although diffusion models have been applied in several fields, but their application expensive in practice, computationally expensive, and time-consuming. Therefore, their application in the real world is limited.

DDPMs have demonstrated outstanding sample quality and variety, but their pricey sampling prevents them from being used just yet. In order to make GANs and DDPMs more practical, we propose DiffGAN-VC, a new unsupervised non-parallel many-to-many voice conversion (VC) method, which introduces multimodal conditional GANs for modeling, and reconstruct denoising diffusion probabilistic model. In the experiments, the objective is to enhance the efficiency of the denoising diffusion probabilistic model while simultaneously reducing its computational complexity. We use a large-step denoising operation, thereby reducing the total number of denoising steps and computation. Finally, let denoising diffusion GAN obtain the same sample quality and diversity as the original diffusion model, but also have the characteristics of high speed and low computational cost of the original GANs. In this paper, we use CycleGAN as our baseline model, based on which we introduce the DDPM and reconstruct the GANs, and also use an expressive multimodal distribution to parameterize the denoising distribution to achieve large step size denoising. 

\section{Related Work}

\subsection{CycleGAN-VC/VC2}

CycleGAN-VC/VC2 \cite{DBLP:conf/eusipco/KanekoK18,DBLP:conf/icassp/KanekoKTH19} is a voice conversion model consisting of two generators G and two discriminators D. CycleGAN-VC/VC2 emerges as a novel approach in the domain of voice conversion, demonstrating its effectiveness in learning the transformation between different acoustic feature sequences without the need for parallel data. These advancements contribute to the ongoing progress in voice conversion research and its applications in various fields. The primary objective of CycleGAN-VC/VC2, as discussed in the research papers by Kaneko and Kameoka \cite{DBLP:conf/eusipco/KanekoK18,DBLP:conf/icassp/KanekoKTH19}, is to acquire the ability to transform acoustic feature sequences belonging to source domain $X$, into those of target domain $Y$, without the reliance on parallel data. The acoustic feature sequences are represented by ${\rm{x}} \in {R^{Q \times T}}$ and ${\rm{y}} \in {R^{Q \times T}}$, where $Q$ and $T$ represents the feature dimension and the sequence length respectively.

The foundation of CycleGAN-VC/VC2 lies in the inspiration drawn from CycleGAN, an originally proposed technique for image-to-image style transfer in the field of computer vision. By leveraging the principles of CycleGAN, CycleGAN-VC/VC2 aims to learn the mapping function $G(X) \to Y$, which facilitates the conversion of an input $x \in X$ to an output $y \in Y$. To achieve this goal, CycleGAN-VC/VC2 employs several loss functions during the learning process. These include adversarial loss, cyclic consistency loss, and identity mapping loss, which collectively contribute to enhancing the quality and fidelity of the generated outputs. Furthermore, CycleGAN-VC2 \cite{DBLP:conf/icassp/KanekoKTH19} introduces an additional adversarial loss to further refine and improve the fine-grained details of the reconstructed features.

\noindent {\bf Generator:} CycleGAN-VC aims to capture diverse temporal structures while maintaining the input structure's integrity. To accomplish this, the architecture of CycleGAN-VC comprises three essential components: a downsampling layer, a residual layer, and an upsampling layer. These components work collaboratively to effectively capture a broad range of temporal relationships present in the data. By incorporating these design choices, CycleGAN-VC can successfully preserve the original structure while accommodating a wide variety of temporal variations.

In the case of CycleGAN-VC2 \cite{DBLP:conf/icassp/KanekoKTH19}, a 2-1-2D CNN network architecture is introduced as an extension to CycleGAN-VC. This modified architecture leverages the advantages of both 2D and 1D CNNs to enhance the performance of the voice conversion system. Specifically, 2D CNNs are employed in the upsampling and downsampling blocks, allowing the model to capture spatial and temporal dependencies simultaneously. This capability facilitates the extraction of more comprehensive and meaningful representations from the input data. On the other hand, 1D CNNs are utilized in the remaining blocks of the network, enabling the model to focus on capturing local temporal relationships and preserving the fine-grained details of the acoustic features. By combining these two types of CNNs in a carefully designed network architecture, CycleGAN-VC2 aims to improve the voice conversion performance by effectively capturing both global and local temporal structures.

The integration of a 2-1-2D CNN network architecture in CycleGAN-VC2 demonstrates a significant advancement in voice conversion research. This innovative design choice allows the model to exploit the benefits of both 2D and 1D CNNs, enabling more efficient and effective processing of temporal information in the voice conversion process. By capitalizing on the strengths of each CNN variant, CycleGAN-VC2 showcases its potential to achieve improved performance in capturing a wide range of temporal structures while preserving the input structure and maintaining the fidelity of the converted acoustic features. These advancements contribute to the ongoing progress in the field of voice conversion and hold promise for various applications that rely on accurate and high-quality voice transformation.

\noindent {\bf Discriminator}: CycleGAN-VC, as introduced in the work by Kaneko \cite{DBLP:conf/eusipco/KanekoK18}, incorporates a discriminator structure based on a 2D convolutional neural network (CNN). This architectural choice enables the discriminator to effectively discriminate data by analyzing the 2D spectral textures present in the input. The utilization of a fully connected layer in the final layer of the discriminator further enhances its discriminative capability by considering the overall input structure. By combining these components, CycleGAN-VC successfully discriminates and distinguishes between different data samples, facilitating the voice conversion process.

However, one challenge associated with employing a 2D CNN structure in the discriminator is the potential increase in the number of parameters, which can negatively impact the model's efficiency and computational requirements. To address this issue, CycleGAN-VC2, introduced by Kaneko et al. \cite{DBLP:conf/icassp/KanekoKTH19}, introduces a technique called Patch-GAN. Patch-GAN modifies the discriminator architecture by incorporating convolution in the final layer, the Patch-GAN approach offers a more parameter-efficient solution while still maintaining the discriminative capability of the discriminator. This modification enables CycleGAN-VC2 to mitigate the issue of a large number of parameters, thereby improving its efficiency and reducing computational complexity.

The introduction of Patch-GAN in CycleGAN-VC2 represents a significant advancement in voice conversion research. This modification not only addresses the challenge of parameter efficiency but also ensures the model's ability to discriminate and differentiate between input samples effectively. By leveraging convolutional layers instead of fully connected layers, CycleGAN-VC2 achieves a more streamlined and efficient architecture, facilitating faster training and inference while maintaining high discriminative performance. 

\subsection{DDPM}
\begin{figure}[!t]
    \centering
    \includegraphics[width=0.9\linewidth]{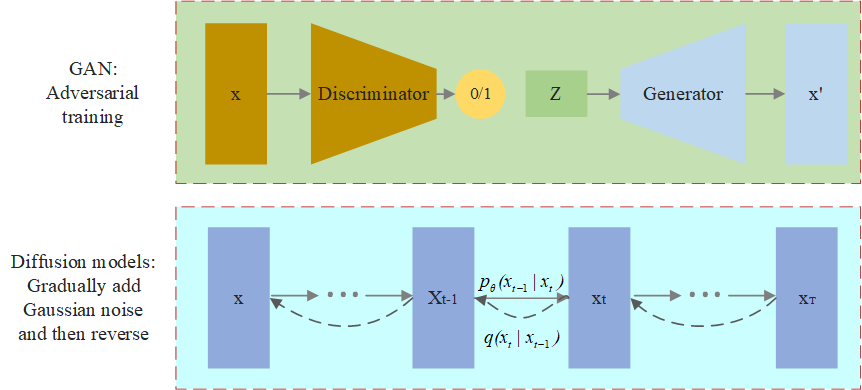}
    \caption{ The distinction between DDPM and GANs is depicted. The preceding illustration illustrates the adversarial training procedure employed by GANs, whereas the subsequent depiction showcases the gradual introduction and removal of Gaussian noise in the diffusion process.}
    \label{GAN_Diff}
\end{figure}

DDPM \cite{DBLP:conf/nips/HoJA20, DBLP:conf/icassp/LuWWRYT22} has good performance in various generative tasks such as image generation \cite{DBLP:conf/cvpr/LugmayrDRYTG22, DBLP:conf/siggraph/SahariaCCLHSF022}, neural acoustic coding \cite{DBLP:conf/iclr/KongPHZC21, DBLP:conf/iclr/ChenZZWNC21}, speech enhancement\cite{DBLP:conf/icassp/LuWWRYT22}, and speech synthesis \cite{DBLP:conf/aaai/Liu00CZ22, DBLP:conf/interspeech/JeongKCCK21}. DDPM utilizes Markov chains to gradually transform simple distributions with Gaussian distributions into complex data distributions. In order to learn the transformation model, there are two distinct processes at play in DDPM: the forward diffusion process and the reverse generation process. The forward diffusion process involves a gradual incorporation of Gaussian noise into the data, leading to its transformation into random noise over time. This process serves to introduce controlled perturbations that explore the data distribution. On the other hand, the reverse generation process focuses on denoising, aiming to restore the original data by mitigating the impact of the noise accumulated during the diffusion process. Consequently, the diffusion process plays a crucial role in effectively denoising the data, facilitating the recovery of its inherent characteristics. Then the reverse process is a denoising process. Gradually denoising can generate a real sample, so the reverse process is also the process of generating data process. DDPM avoids the generator and discriminator in GANs being less stable during training and the "backward collapse" problem. Compared with GANs, diffusion model training is more stable and can generate more diverse high-quality data. However, since DDPM requires hundreds of iterations to generate data, their inference speed is relatively slow compared with VAEs and GANs. The comparison between diffusion model and GAN is shown in Figure \ref{GAN_Diff}. The current development of DDPM in speech translation is hampered by two main challenges: (1) Although DDPM is essentially gradient-based model, the guarantee of high sample quality is usually at the expense of thousands or hundreds of denoising steps. This limits the wide application of DDPM in real life. (2) When reducing the denoising step, the diffusion model exhibits significant degradation in model convergence due to the complex data distribution, resulting in blurry and over-smoothed predictions in the mel-spectrogram.

\section{Method}
Generative Adversarial Networks (GANs) can quickly generate high-quality samples, but suffer from poor pattern coverage and lack of diversity in the generated samples. Although the diffusion model beats GANs in image generation, the inverse process is computationally expensive. To speed up the diffusion model and reduce computation, we use a large-step denoising operation, thereby reducing the total number and computation of denoising steps. Meanwhile, we introduce multimodal conditional GANs in the diffusion model, developing a new generative model, which we call denoising diffusion GANs. This paper adopts CycleGAN as the baseline model, and proposes a novel reconstruction method for the denoising diffusion model. At the same time, we also use an expressive multimodal distribution to parameterize the denoising distribution to achieve large stride denoising and improve model performance. Below we introduce the large-step denoising operation, the multimodal distribution design, and the parameterization of the denoising distribution, respectively.

The forward process of diffusion model \cite{DBLP:conf/nips/HoJA20} which gradually adds Gaussian noise to the data ${x_0} \to q({x_0})$ is performed over a discrete T time step. Where $q({x_0})$ is the data distribution of a true Mel-spectrogram, and ${x_0}$ denotes a sample from input data $ {X} $. ${x_t}$ is sampled from a unit Gaussian independent from $x_0$ using the Markovian forward process as follows:
\begin{equation}
q({x_{1:T}}|{x_0}) = \mathop \prod \limits_{t \ge 1} \mathbb{N} ({x_t};\sqrt {1 - {\beta _t}} {x_{t - 1}},{\beta _t}I)
\end{equation}
 where, ${t \in [1:T]}$, and ${\beta _{\rm{t}}}$ is pre-defined variance schedule. The denoising process \cite{DBLP:conf/nips/HoJA20} is defined as:
\begin{equation}
{p_\theta }({x_{0:T}}) = p({x_T})\mathop \prod \limits_{t \ge 1} \mathbb{N} ({x_{t - 1}};{\mu _\theta }({x_t},t),\sigma _t^2I)
\end{equation}
where ${\sigma _t^2}$ is fixed according to each ${t}$. 
DDPMs typically presume that the denoising distribution may be roughly modeled as a Gaussian distribution. It is well-established that the assumption of Gaussianity is only applicable in the context of infinitesimally small denoising steps. This limitation results in a slow sampling problem when a large number of denoising steps are employed in the reverse process. In order to speed up the model operation and reduce the amount of computation, we optimize the model by reducing the number of denoising steps T. 
% (and reducing the number of denoising diffusion steps T required in the reverse process of the diffusion model).

\begin{figure*}[t]
    \centering
    \includegraphics[width=1.0\linewidth]{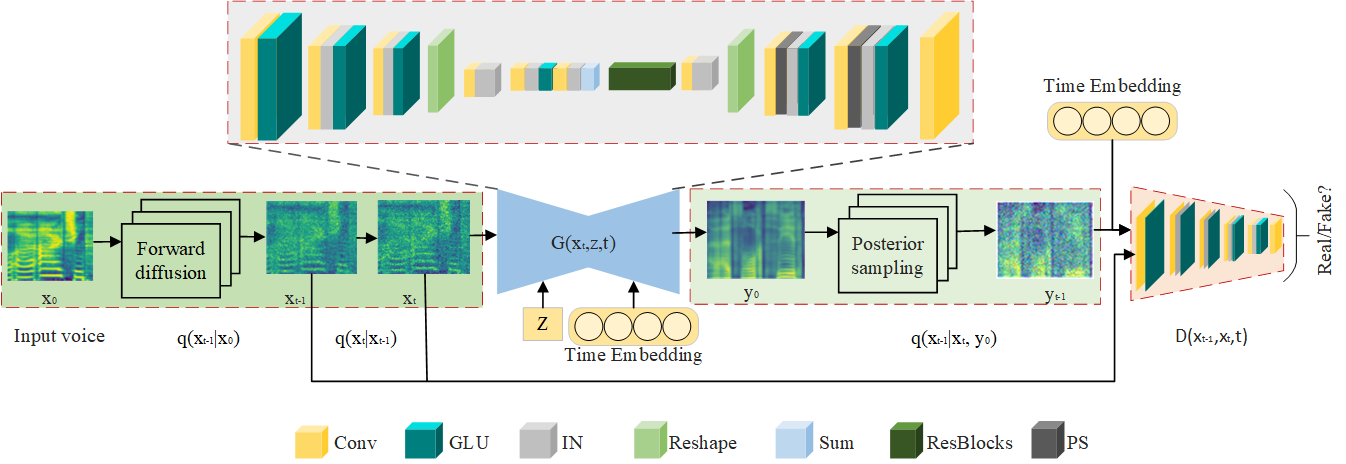}
    \caption{\centering{The overall pipeline of DiffGAN-VC, where ${G(x,z,t)}$ is the generator of our model. ${D(x_{t-1},x_t,t)}$ denotes the discriminator of our model, ${q({x_t}|{x_{t - 1}})}$ is thus the true denoising diffusion transition}}
    \label{DiffGAN-VC}
\end{figure*}

Using a larger step size (that is, fewer denoising steps) in the reverse process will have two problems. Firstly, the assumption of Gaussianity in the denoising distribution cannot be guaranteed to hold under large denoising steps and non-Gaussian data distributions. Furthermore, with each subsequent enhancement of the denoising process, the distribution for noise removal becomes increasingly intricate and exhibits multiple modes. To address this, a non-Gaussian multimodal distribution can be constructed to accurately model the denoising distribution. In light of the ability of conditional GANs to accurately simulate complex conditional distributions within the image domain, we adopt conditional GANs to approximate the true denoising distribution. We use CycleGAN as the base model, then modify this model to conditional GANs, and finally fuse this model with the diffusion model , forming a new denoising diffusion GANs. The overall pipeline of the DiffGAN-VC is depicted in the Figure \ref{DiffGAN-VC}.

Within our framework, we express the training procedure as the alignment between a conditional GAN generator and its ability to execute operations $p_\theta(x_{t - 1}|x_t)$ and $q(x_t|x_{t - 1})$. This objective is accomplished through the utilization of an adversarial loss, aiming to decrease the divergence $D_{adv}$ during every denoising iteration.

% The objective function for training is defined as follows:

% \begin{equation}
% \min_\theta \sum_{t \ge 1} \mathcal{L}_{q(x_t)}[D_{adv}(q(x_{t - 1}|x_t)||p_\theta(x_{t - 1}|x_t))]
% \end{equation}

In order to facilitate the integration of adversarial training into our model, we introduce a discriminator that is dependent on time and denoted as $D_\theta(x_{t - 1},x_t,t): \mathbb{R}^N \times \mathbb{R}^N \times \mathbb{R} \to [0,1]$, where $\theta$ represents the associated parameters. This time-dependent discriminator is designed to assess the plausibility of $x_{t - 1}$ being a denoised version of $x_t$, taking into account their respective n-dimensional inputs. During the training process, the discriminator is optimized by following a specific objective. The aim is to bolster the discriminator's capacity to differentiate between the denoised and noisy variants of the input data, thereby ultimately enhancing its discriminatory prowess.

\begin{align}
    \label{eq:objective}
    \min_\varphi \sum_{t \ge 1} &\mathcal{L}{q(x_t)}\mathcal{L}{q(x_{t - 1}|x_t)}[-\log D_\varphi(x_{t - 1},x_t,t)] \nonumber \\
    &+ \mathcal{L}{p\theta(x_{t - 1}|x_t)}[-\log(1 - D_\varphi(x_{t - 1},x_t,t))]
    \end{align}

    Our approach involves the comparison between synthetic samples derived from the generator $p_\theta(x_{t - 1}|x_t)$ and authentic samples obtained from the conditional distribution $q(x_t|x_{t - 1})$. It is worth highlighting that the generation of the first expectation necessitates the sampling of data points from a distribution, namely $q(x_t|x_{t - 1})$, which is not explicitly known to us. This introduces an additional challenge in effectively estimating and evaluating the expected values.

Given the discriminator, we train the generator using the following objective:

\begin{equation}
\max_\theta \sum_{t \ge 1} \mathcal{L}_{q(x_t)p_\theta(x_{t - 1}|x_t)}[\log(D_\varphi(x_{t - 1},x_t,t))]
\end{equation}

In order to achieve the denoising step, the objective of our model involves updating the generator using a non-saturating GAN objective.

The diffusion model introduces a parameterized approach through ${f_\theta}({x_{t - 1}}|{x_t})$. This means that the denoising model incorporates a parameterization scheme, allowing for the estimation of ${x_{t - 1}}$ based on ${x_t}$. Initially, the denoising model ${p_\theta}({x_{t - 1}}|{x_t})$ predicts the value of ${x_0}$. Subsequently, ${x_{t - 1}}$ is sampled from the posterior distribution $q({x_{t - 1}}|{x_t},{x_0})$, taking into account both ${x_t}$ and the predicted ${x_0}$.

The distribution ${q({x_{t - 1}}|{x_0},{x_t})}$ provides an intuitive representation of the distribution of ${x_{t - 1}}$ during the denoising process, which occurs from ${x_t}$ to ${x_0}$. Similarly, we define ${p_\theta}({x_{t - 1}}|{x_t})$ in the following manner:

\begin{equation}
{p_\theta}({x_{t - 1}}|{x_t}) = {G_\theta}({x_t},z,t))dz
\end{equation}

The distribution ${p_\theta}({x_0}|{x_t})$ captures the implicit nature of the GAN generator ${G_\theta}({x_t},z,t)$, which is conditioned on both ${x_t}$ and the L-dimensional latent variable ${z}$. This generator is responsible for producing the unperturbed version ${x_0}$. However, in our specific scenario, the role of the generator is to solely predict the unperturbed ${x_0}$ and subsequently reintroduce the perturbation through the utilization of $q({x_{t - 1}}|{x_t},{x_0})$. This enables us to maintain the perturbation information while generating ${x_{t - 1}}$ based on the observed ${x_t}$ and the predicted unperturbed ${x_0}$.

% GANs are known to suffer from training instability and mode collapse, some possible causes include difficulty generating samples directly from complex distributions in one go, and overfitting problems when the discriminator only looks at clean samples. In contrast, our model decomposes the generation process into several conditional denoising diffusion steps, each of which is relatively simple to model due to the strong conditioning on ${x}$. Also, the diffusion process smoothes the data distribution, making the discriminator less likely to overfit. Therefore, we expect our model to exhibit better training stability and pattern coverage.
GANs are known to have mode collapse and training instability. The potential causes include the inability to generate samples directly from complicated distributions in a single step and overfitting issues when the discriminator only considers clean samples. In contrast, due to the strong conditioning on ${x}$, our model divides the generative process into multiple stages of conditional denoising steps, each of which is very simple to model. Furthermore, the diffusion process smoothes the data distribution and alleviates the overfitting problem in the discriminator. Therefore, our model will have higher pattern coverage and training stability.
DDPM avoids the generator and discriminator in GAN being less stable during training and the "backward collapse" problem. Compared with GAN, diffusion model training is more stable and can generate more diverse high-quality data. However, since DPM requires hundreds of iterations to generate data, their inference speed is relatively slow.

% we apply the diffusion model to the CycleGAN model and then minimize the corresponding adversarial loss function.
% GANs are known to suffer from training instability and mode collapse, some possible causes include difficulty generating samples directly from complex distributions in one go, and overfitting problems when the discriminator only looks at clean samples. In contrast, our model decomposes the generation process into several conditional denoising diffusion steps, each of which is relatively simple to model due to the strong conditioning on ${x}$. Also, the diffusion process smoothes the data distribution, making the discriminator less likely to overfit. Therefore, we expect our model to exhibit better training stability and pattern coverage.

\section{Experiments}
\subsection{Experimental Setup}
{\bf Dataset and Conversion Process:}
We evaluate our method objectively and subjectively on part of the Speech Conversion Challenge (VCC) 2020 data set, which is a semi-parallel speech data set containing input and output speech from two women and two men, each 70 English sentences, 20 parallel sentences, and 50 non-parallel sentences. There are a total of 4 × 3 = 12 distinct source-target combinations that can be derived. The number of sentences used for training, and test are 70 and 25, respectively. In our experiments, all speech signals are sampled at 16 kHz. From every sentence, we utilized the WORLD toolkit to extract various acoustic features, including 35 mel-cepstral coefficients (MCEPs), logarithmic fundamental frequencies ($\log {F_0}$), and aperiodicities (APs). These features provide valuable representations of the speech signal, capturing essential characteristics related to spectral information, pitch contour, and the irregular components of the signal. The extraction process involved careful analysis and computation to ensure accurate and informative representations of the speech data. We apply the DiffGAN-VC to MCEP transformation, for $\log {F_0}$, we use log-Gaussian normalized transformation for transformation, use Aps directly, and finally use WORLD vocoder to synthesize speech.
% \begin{table*}[htp]
% %   \setlength{\abovecaptionskip}{0pt}%    
% %   \setlength{\belowcaptionskip}{10pt}%
%   \caption{Mean and standard error with different models}
%   \centering
%   \fontsize{8.7}{7}\selectfont
%   \label{Comparison}
%     \begin{tabular}{|l|c|c|}
%     \toprule
%     {Methods}& MCD ${\downarrow}$ & MOS ${\uparrow}$\\
%     \midrule
%     VQVC & 6.08  & 2.86 $\pm$ 0.78 \cr
%     StarGAN-VC2 & 5.04  & 3.64 $\pm$ 0.52 \cr
%     CycleGAN-VC2& 4.87  & 2.96 $\pm$ 0.50 \cr
%      DDPM & 4.78  & 3.71 $\pm$ 0.49  \cr
%     \midrule
%     \textbf{Our model} &\textbf{4.61 } & \textbf{3.96 $\pm$ 0.51}\cr
%     \bottomrule
%     \end{tabular}
% \end{table*}
\begin{table*}[htp]
  \caption{Mean and standard error with different models}
  \centering
  \fontsize{8.7}{7}\selectfont
  \label{Comparison}
    \begin{tabular}{lcccc}
    \toprule
    \multirow{2}{*}{\textbf{Methods}}&
    \multicolumn{2}{c}{\textbf{intra-gender}}&\multicolumn{2}{c}{\textbf{ inter-gender}}\cr
    \cmidrule(lr){2-3} \cmidrule(lr){4-5}
    & MCD ${\downarrow}$ & MOS ${\uparrow}$ & MCD ${\downarrow}$ & MOS ${\uparrow}$\cr
    \midrule
    VQVC 
    & 6.41  & 3.26 $\pm$ 0.36 
    & 6.71  & 2.89 $\pm$ 0.37 \cr
    StarGAN-VC2  
    & 6.45  & 3.36 $\pm$ 0.42 
    & 5.85  & 3.35 $\pm$ 0.53 \cr
    CycleGAN-VC2
    & 6.09  & 3.71 $\pm$ 0.45 
    & 5.78 & 2.93 $\pm$ 0.47 \cr
    DDPM 
    & 5.45  & 3.47 $\pm$ 0.43 
    & 5.76  & 3.47 $\pm$ 0.57 \cr
    \midrule
    \textbf{Our model} &\textbf{5.99} & \textbf{3.75 $\pm$ 0.42} 
    & \textbf{5.62 } &  \textbf{3.86 $\pm$ 0.51}\cr
    \bottomrule
    \end{tabular}
\end{table*}

\noindent {\bf Experimental details:}
We design the architecture based on CycleGAN-VC2, employing a 2-1-2D CNN in the generator (G) and a 2D CNN in the discriminator (D). During the experiments, the input to the network is adjusted to ${x_t}$, and temporal embedding is utilized to ensure conditioning on the time step (t). The latent variable (z) is responsible for controlling the normalization layer, where a multi-layer FC network predicts the displacement and scale parameters in the group normalization. The training procedure encompassed a total of $2 \times {10^5}$ iterations. Moreover, to enhance the optimization process, a momentum term of 0.5 was incorporated, aiding in the stabilization and convergence of the training procedure.

In the conducted experiments, we categorized the 12 different combinations into two groups: inter-gender conversion and intra-gender conversion. The inter-gender conversion includes transformations from female to male and from male to female, while the intra-gender conversion includes transformations within the same gender, specifically from female to female and from male to male.

\subsection{Objective Evaluation}

We chose the VQVC, StarGAN, CycleGAN and DDPM based methods as the comparison for our experiments, and objectively evaluated the results to verify that the DiffGAN-VC has a certain improvement and optimization in model performance. To conduct a thorough analysis and assess the performance of the models, we employed the MCD metric, which serves as a measure of global structural dissimilarities. The MCD metric calculates the distance between the log-modulated spectra of the target and converted MCEPs. A smaller MCD value indicates a higher level of similarity between the target and transformed features, reflecting a better quality of conversion. In our evaluation, we generated a total of 300 sentences, obtained through 4 × 3 source-target combinations. Each combination consisted of 25 sentences. This comprehensive set of sentences allowed us to evaluate across various source-target pairs and assess the effectiveness of the conversion process. We compare DiffGAN-VC with VQVC, StarGAN-VC2, CycleGAN-VC2 and DDPM, which are listed in Table \ref{Comparison}, respectively. In Table \ref{Comparison}, DiffGAN-VC outperforms other models in terms of MCD. This suggests that the introduction of multimodal diffusion methods in conditional GANs models is useful for improving feature quality. These experiments confirm that the proposed method effectively reduces the distance between the transformed acoustic feature sequence and the target sequence.

% We invited multiple English-educated testers to conduct listening tests, and compared the performance of DiffGAN-VC, DiffGAN-VC performed well in multi-domain non-parallel VC. To measure naturalness, we conducted a Mean Opinion Score (MOS) test (5: good to 1: bad), with higher scores indicating better naturalness. We randomly select 32 sentences longer than 2 seconds and shorter than 5 seconds from the evaluation set. The results are shown in Table \ref{Comparison}. The other is the Voice Similarity Score (VSS) test, which measures how similar the timbre of the transformed voice is to the timbre of the ground truth. And its scoring mechanism is the same as MOS, the tester chooses a score from the similarity of the converted voice on a scale of 1-5, the higher the score, the more similar the timbre. We randomly selected 12 sample pairs from the evaluation set. Thirteen well-educated English speakers participated in the test. The results are shown in Figure \ref{VSS}, where S and T represent source and target speakers, respectively, and F and M represent female and male speakers, respectively. Table \ref{Comparison} and Figure \ref{VSS} show the naturalness of the MOS and the similarity VSS of the source and target speakers, respectively. 
A comprehensive evaluation was conducted to assess the performance of the DiffGAN-VC method in multi-domain non-parallel voice conversion. The evaluation involved multiple English-educated testers, and several listening tests were carried out. The primary focus was on evaluating naturalness and voice similarity.

To measure naturalness, a MOS test was employed. Testers assigned scores ranging from 5 (good) to 1 (bad), with higher scores indicating better naturalness. A subset of 32 sentences, with durations between 2 and 5 seconds, was randomly selected from the evaluation set for this test. The obtained results, including MOS scores, are presented in Table \ref{Comparison}.

\begin{figure}[t]
    \centering
    \includegraphics[width=0.7\linewidth]{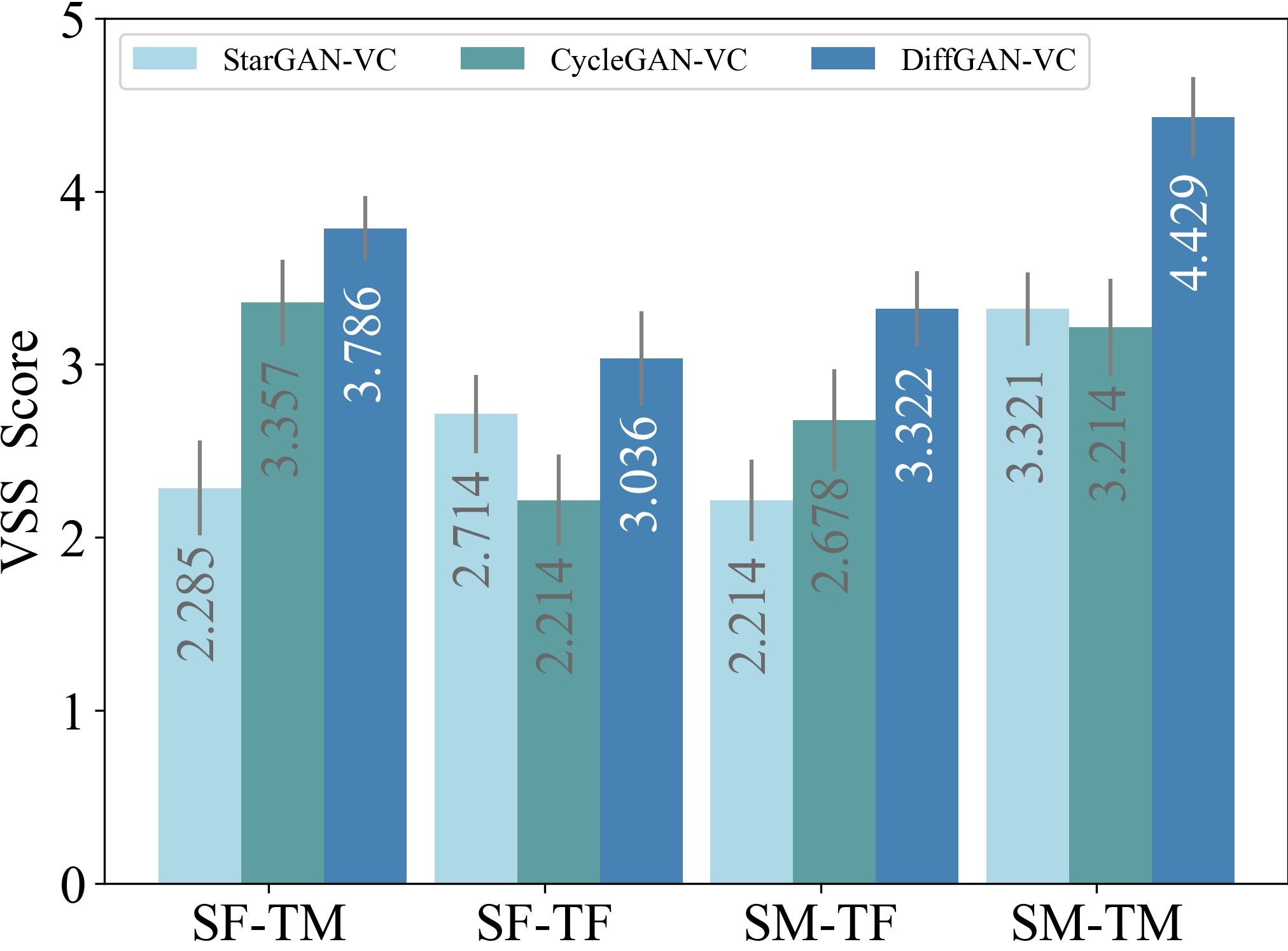}
    \caption{VSS for speaker similarity}
    \label{VSS}
\end{figure}
\subsection{Subjective Evaluation}

Additionally, we conduct Voice Similarity Score (VSS) evaluation to assess the similarity of the transformed voices to the original ones. The scoring system used in the VSS test was the same as the MOS test, with testers assigning scores from 1 to 5 based on perceived similarity. We randomly select 12 sample pairs for this test, which involved thirteen well-educated English speakers. Figure \ref{VSS} illustrates the outcome of the VSS evaluation, in which the labels T and S denote the target and source speakers, respectively. Female and male speakers are represented by F and M respectively. The MOS naturalness and VSS similarity scores for the source and target speakers are presented in Table \ref{Comparison} and Figure \ref{VSS}, respectively.

The evaluation results revealed several significant findings. Firstly, the DiffGAN-VC method demonstrated a substantial improvement over the baseline model in both inter-gender and intra-gender voice conversion tasks. Moreover, when considering the overall performance, DiffGAN-VC exhibited significant advancements compared to the baseline CycleGAN-VC, both in terms of naturalness and speaker similarity.

\section{Conclusion}
The primary goal of this research is to enhance the performance of Generative Adversarial Networks (GANs) and Denoising Diffusion Probabilistic Models (DDPM) in various aspects, including speed, diversity, utilization of non-parallel training data, and the coverage and diversity of generated samples. To accomplish this, we have made modifications to the existing CycleGAN-VC2 system by incorporating the embedding of a stride denoised multimodal diffusion model. Both the generator and discriminator are conditioned on this embedding, facilitating the targeting of specific source and target speakers.  DiffGAN-VC has been thoroughly evaluated using a limited training dataset. The evaluation demonstrates that DiffGAN-VC outperforms CycleGAN-VC in terms of both objective and subjective metrics. In this study, we have applied the DiffGAN-VC model to speaker identity speech conversion. 
It is important to note that this approach can be expanded to encompass additional tasks, such as multi-emotional voice conversion and music genre conversion. Exploring these additional applications represents a compelling direction for future research in this domain. 

\section{Acknowledgement}
This paper is supported by the Key Research and Development Program of Guangdong Province under grant No.2021B0101400003. Corresponding author is Jianzong Wang from Ping An Technology (Shenzhen) Co., Ltd (jzwang@188.com).

\bibliographystyle{splncs04}
\bibliography{refs,0-citation-self}
\end{document}